 \documentclass[final,3p,times]{elsarticle}

\usepackage{multirow,setspace,amssymb,amsmath,graphicx,color,rotating,subfigure,url,lineno,natbib}
\usepackage{textcomp}

\journal{Physics Letters A} 

\begin{document}

\begin{frontmatter}

\title{Degree distributions of the visibility graphs mapped from fractional Brownian motions and multifractal random walks}%
\author[BS,SS,RCE]{Xiao-Hui Ni}
\author[BS,SS,RCE,ETH]{Zhi-Qiang Jiang}
\author[BS,SS,RCE,RCSE,RCFE]{Wei-Xing Zhou \corref{cor}}
\cortext[cor]{Corresponding author. Address: 130 Meilong Road, P.O.
Box 114, School of Business, East China University of Science and
Technology, Shanghai 200237, China, Phone: +86 21 64253634, Fax: +86
21 64253152.}
\ead{wxzhou@ecust.edu.cn} %

\address[BS]{School of Business, East China University of Science and Technology, Shanghai 200237, China}
\address[SS]{School of Science, East China University of Science and Technology, Shanghai 200237, China}
\address[RCE]{Research Center for Econophysics, East China University of Science and Technology, Shanghai 200237, China}
\address[ETH]{Chair of Entrepreneurial Risks, D-MTEC, ETH Zurich, Kreuplatz 5, CH-8032 Zurich, Switzerland}%
\address[RCSE]{Engineering Research Center of Process Systems Engineering (Ministry of Education), East China University of Science and Technology, Shanghai 200237, China}
\address[RCFE]{Research Center on Fictitious Economics \& Data Science, Chinese Academy of Sciences, Beijing 100080, China}

\begin{abstract}
The dynamics of a complex system is usually recorded in the form of
time series, which can be studied through its visibility graph from
a complex network perspective. We investigate the visibility graphs
extracted from fractional Brownian motions and multifractal random
walks, and find that the degree distributions exhibit power-law
behaviors, in which the power-law exponent $\alpha$ is a linear
function of the Hurst index $H$ of the time series. We also find
that the degree distribution of the visibility graph is mainly
determined by the temporal correlation of the original time series
with minor influence from the possible multifractal nature. As an
example, we study the visibility graphs constructed from two Chinese
stock market indexes and unveil that the degree distributions have
power-law tails, where the tail exponents of the visibility graphs
and the Hurst indexes of the indexes are close to the $\alpha\sim H$
linear relationship.
\end{abstract}


\end{frontmatter}

\section{Introduction}
\label{S1:Intro}

Complex systems are ubiquitous in natural and social sciences, where
the constituents interact with one another and form a complex
network. In recent years, complex network theory has stimulated
explosive interests in the study of social, informational,
technological, and biological systems, resulting in a deeper
understanding of complex systems
\cite{Albert-Barabasi-2002-RMP,Newman-2003-SIAMR,Dorogovtsev-Mendes-2003,Boccaletti-Latora-Moreno-Chavez-Hwang-2006-PR}.
However, for many complex systems, it is hard to obtain detailed
information of interacting constituents and their ties, which makes
the underlying network invisible. Instead, we are able to observe
and record a time series generated by the system. For such cases,
time series analysis becomes a crucial way to unveil the dynamics of
complex systems. There are also some efforts to map time series into
graphs to study time series from the network perspective, which
amounts to investigating the dynamics from the associated network
topology.

For a pseudoperiodic time series, one can partition it into disjoint
cycles according to the local minima or maxima, and each cycle is
considered a basic node of a network, in which two nodes are deemed
connected if the phase space distance or the correlation coefficient
between the corresponding cycles is less than a predetermined
threshold \cite{Zhang-Small-2006-PRL}. We note that a weighted
network can also be constructed if the phase space distance or the
correlation coefficient is treated as the weight of a link. This
method for pseudoperiodic time series can also be generalized to
other time series, where a node is defined by a sub-series of a
fixed length as the counterpart of a cycle, which has been applied
to stock prices \cite{Yang-Yang-2008-PA}.

Another method for network construction from time series is based on
the fluctuation patterns \cite{Li-Wang-2006-CSB,Li-Wang-2007-PA}. In
this approach, each data point is encoded as a symbol {\textbf{R}},
{\textbf{r}}, {\textbf{D}}, or {\textbf{d}}, corresponding to big
rise, small rise, big drop and small drop, respectively. The time
series is then transformed into a symbol sequence. Defining a
fluctuation pattern as an $n$-tuple consist of a string of $n$
symbols, the symbol sequence can be further mapped into a sequence
of non-overlapping $n$-tuples. The $n$-tuples are treated as nodes
of the constructed network. Therefore, the number of nodes does not
exceed $n^4$. Two nodes are connected if the associated $n$-tuples
appear one after the other in the pattern sequence. Furthermore, the
edge weight between two nodes can be defined as the occurrence
number of two successive patterns in the sequence. This approach has
been applied to study the price trajectory of Hang Seng index
\cite{Li-Wang-2006-CSB,Li-Wang-2007-PA}.

\begin{figure}[htb]
\centering
\includegraphics[width=7cm]{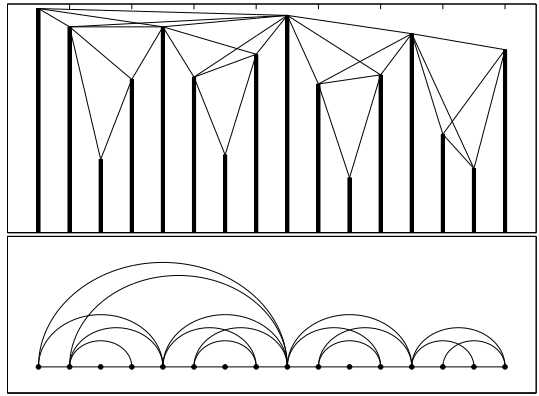}
\caption{\label{Fig:VG_plot} Example of a time series containing 16
data points (upper panel) and the associated visibility graph
derived from the visibility algorithm (lower panel).}
\end{figure}

A third method is to convert time series into visibility graphs
based on the visibility of nodes
\cite{Lacasa-Luque-Ballesteros-Luque-Nuno-2008-PNAS}. For
simplicity, consider an evenly sampled time series $\{y_t:
t=1,2,\cdots,N\}$. Each data point of the time series is encoded
into a node of the visibility graph. Two arbitrary data points $y_i$
and $y_j$ have visibility if any other data point $y_k$ located
between them fulfills
\begin{equation}
 \frac{y_j - y_k}{j-k} > \frac{y_j-y_i}{j-i}~.
\label{Eq:Visibility}
\end{equation}
Two visible nodes become connected in the associated graph. An
example of a time series containing 16 data points and the
associated visibility graph derived from the visibility algorithm is
illustrated in Fig.~\ref{Fig:VG_plot}. By definition, any visibility
graph extracted from a time series is always connected since each
node sees at least its nearest neighbor(s) and the degree of any
node $y_t$ with $1<t<N$ is not less than 2. In addition, a periodic
time series converts into a regular graphs, whose degree
distribution is formed by a finite number of peaks related to the
series period, while random time series lead to irregular random
graphs \cite{Lacasa-Luque-Ballesteros-Luque-Nuno-2008-PNAS}. It is
also found that visibility graph is invariant under affine
transformations of the series data since the visibility criterion is
invariant under rescaling of both horizontal and vertical axes, and
under horizontal and vertical translations
\cite{Lacasa-Luque-Ballesteros-Luque-Nuno-2008-PNAS}.

Degree distribution $p(k)$ is one of the most important
characteristic properties of complex networks
\cite{Barabasi-Albert-1999-Science}. The degree distribution of the
visibility graphs of several specific examples of time series have
been investigated
\cite{Lacasa-Luque-Ballesteros-Luque-Nuno-2008-PNAS}. For a random
time series extracted from an uniform distribution in $[0,1]$, the
degree distribution of the visibility graph has an exponential tail
$p(k)\sim e^{-k/k_0}$. Alternatively, the visibility graphs of
Brownian motions and Conway series are scale-free, characterized by
a power-law tail in the degree distribution:
\begin{equation}
 p(k)\sim k^{-\alpha}~,
 \label{Eq:pk:PL}
\end{equation}
where $\alpha=2.00\pm0.01$ for Brownian motions and
$\alpha=1.2\pm0.1$ for Conway series. It is also conjectured that
the temporal correlation of the time series (characterized by its
Hurst index $H$) might have influence on the degree distribution of
its visibility graph
\cite{Lacasa-Luque-Ballesteros-Luque-Nuno-2008-PNAS}. In this work,
we test this projection based on extensive numerical simulations.
Specifically, fractional Brownian motions (FBMs)
\cite{Mandelbrot-Ness-1968-SIAMR} and multifractal random walks
(MRWs) \cite{Bacry-Delour-Muzy-2001-PRE} are synthesized to
investigate the influence of autocorrelation and multifractality on
the degree distribution.

\section{Numerical analysis}
\label{S1:Numerical}

\subsection{Generating FBMs and MRWs}

\begin{figure}[htb]
\centering
\includegraphics[width=7cm]{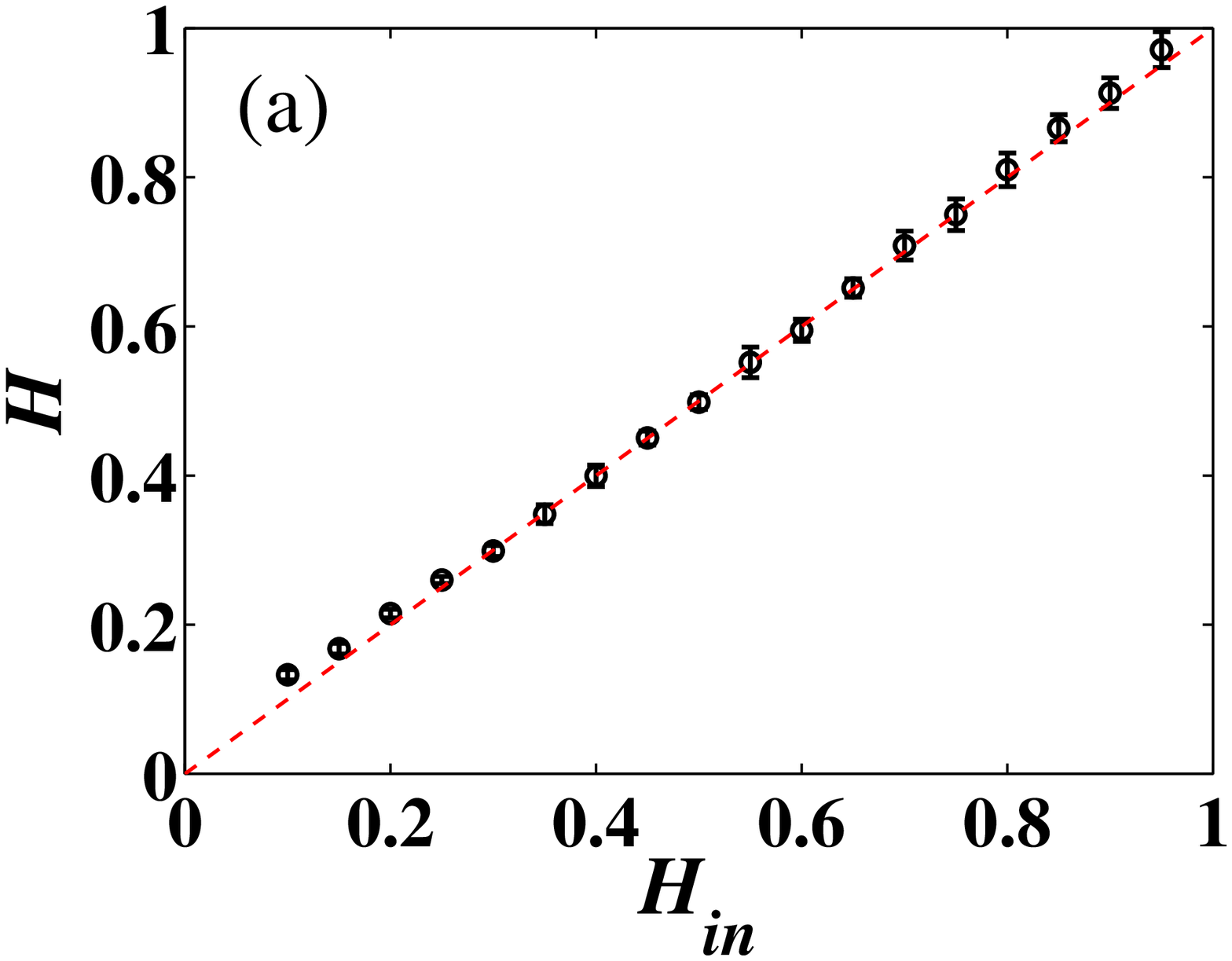}
\includegraphics[width=7cm]{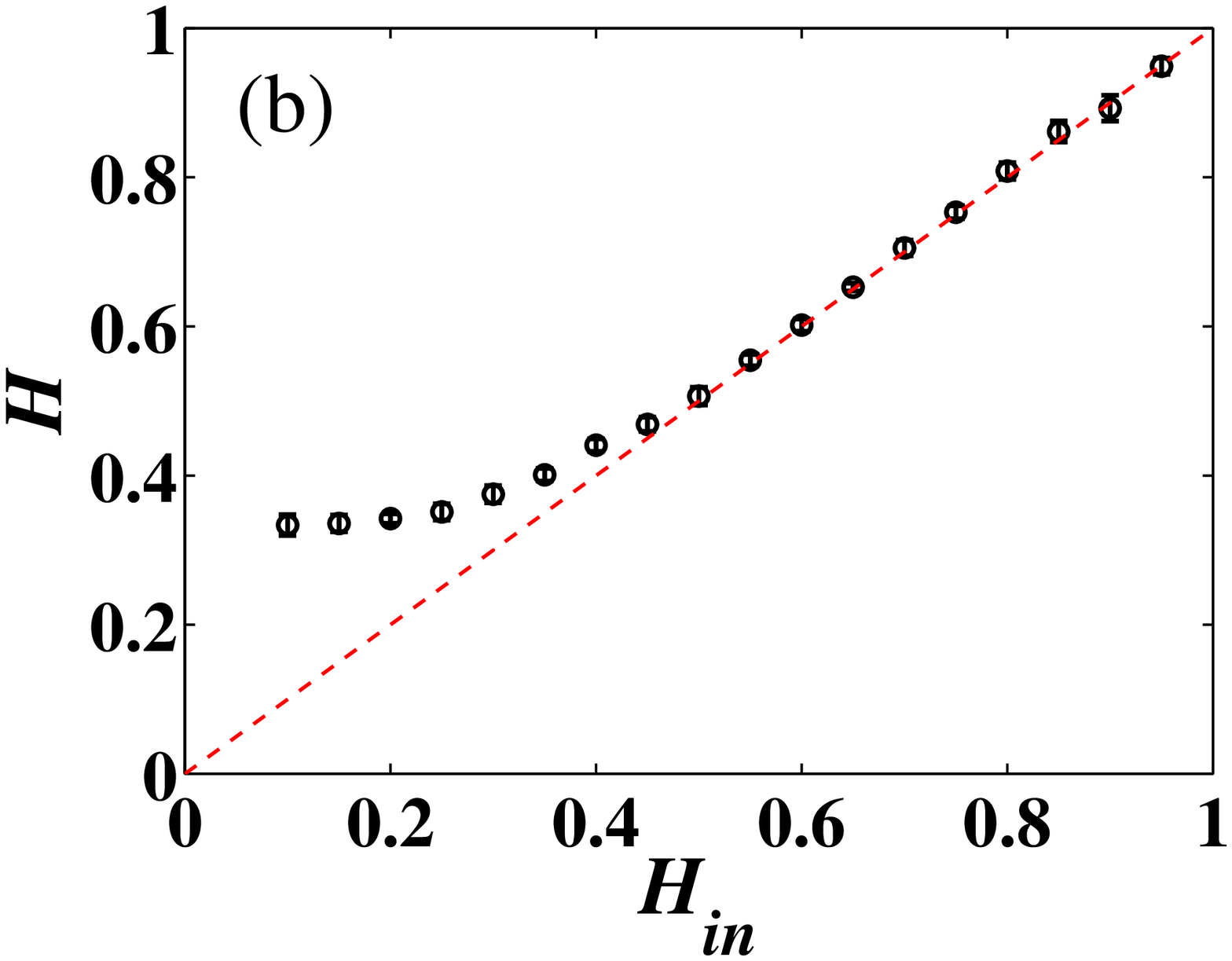}
\caption{\label{Fig:Compare:H} (color online.) Dependence of the
Hurst index $H$ of the simulated (a) FBMs and (b) MRWs, determined
by detrended fluctuation analysis on the input Hurst index
$H_{\rm{in}}$ in the two synthesis algorithms. }
\end{figure}

There are many different algorithms for the generation of fractional
Brownian motions
\cite{Bardet-Lang-Oppenheim-Philippe-Stoev-Taqqu-2003} and we adopt
a wavelet-based algorithm to simulate FBMs
\cite{Abry-Sellan-1996-ACHA}. On the other hand, a multifractal
random walk can be generated by the cumulative summation of the
increments
\begin{equation}
 \Delta y_t = \epsilon_t e^{\omega_t}~,
 \label{Eq:MRW:syn}
\end{equation}
where $\epsilon_t$ is a fractional Gaussian noise with Hurst index
$H_{\rm{in}}$, $\omega_t$ is a correlated Gaussian noise, and they
are independent \cite{Bacry-Delour-Muzy-2001-PRE}. We use the
detrended fluctuation analysis
\cite{Peng-Buldyrev-Havlin-Simons-Stanley-Goldberger-1994-PRE,Kantelhardt-Bunde-Rego-Havlin-Bunde-2001-PA}
to verify if the resultant Hurst index of the generated signals is
identical to the input value of $H_{\rm{in}}$ in the algorithms. For
each $H_{\rm{in}}$, we generate 10 realizations and calculate the
mean Hurst index $H$. The results are presented in
Fig.~\ref{Fig:Compare:H}. The top panel of Fig.~\ref{Fig:Compare:H}
shows that the estimated Hurst indexes of the synthesized FBMs are
very close to the input value $H_{\rm{in}}$ with minor deviation for
small $H_{\rm{in}}$. For MRWs, we find that $H=H_{\rm{in}}$ when
$H_{\rm{in}}\geqslant0.5$ and a systematic deviation $H>H_{\rm{in}}$
when $H_{\rm{in}}\leqslant0.5$. In addition, we have confirmed that
the generated MRW signals possess multifractal nature.

\subsection{Numerical results}

\begin{figure}[htb]
\centering
\includegraphics[width=7cm]{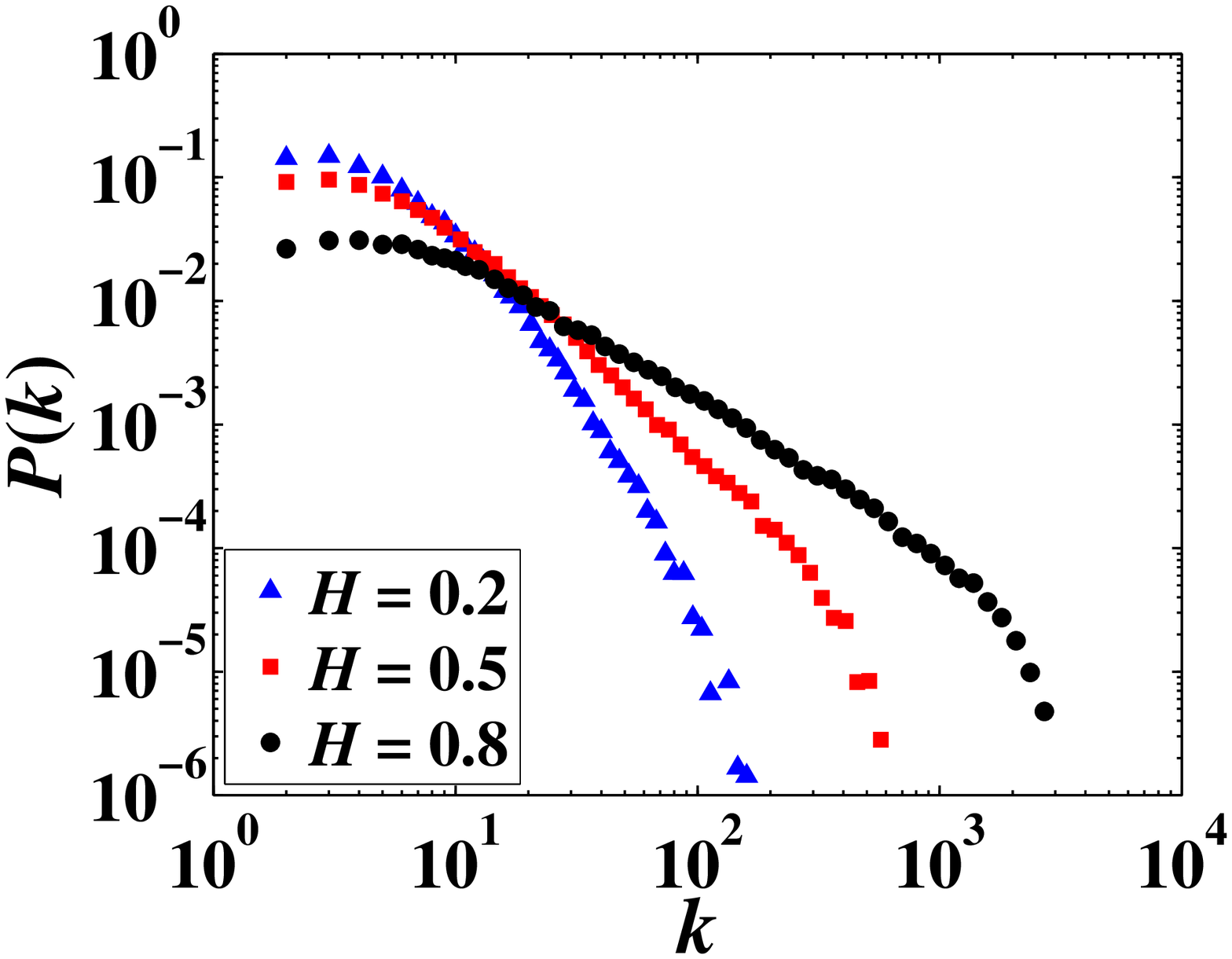}
\includegraphics[width=7cm]{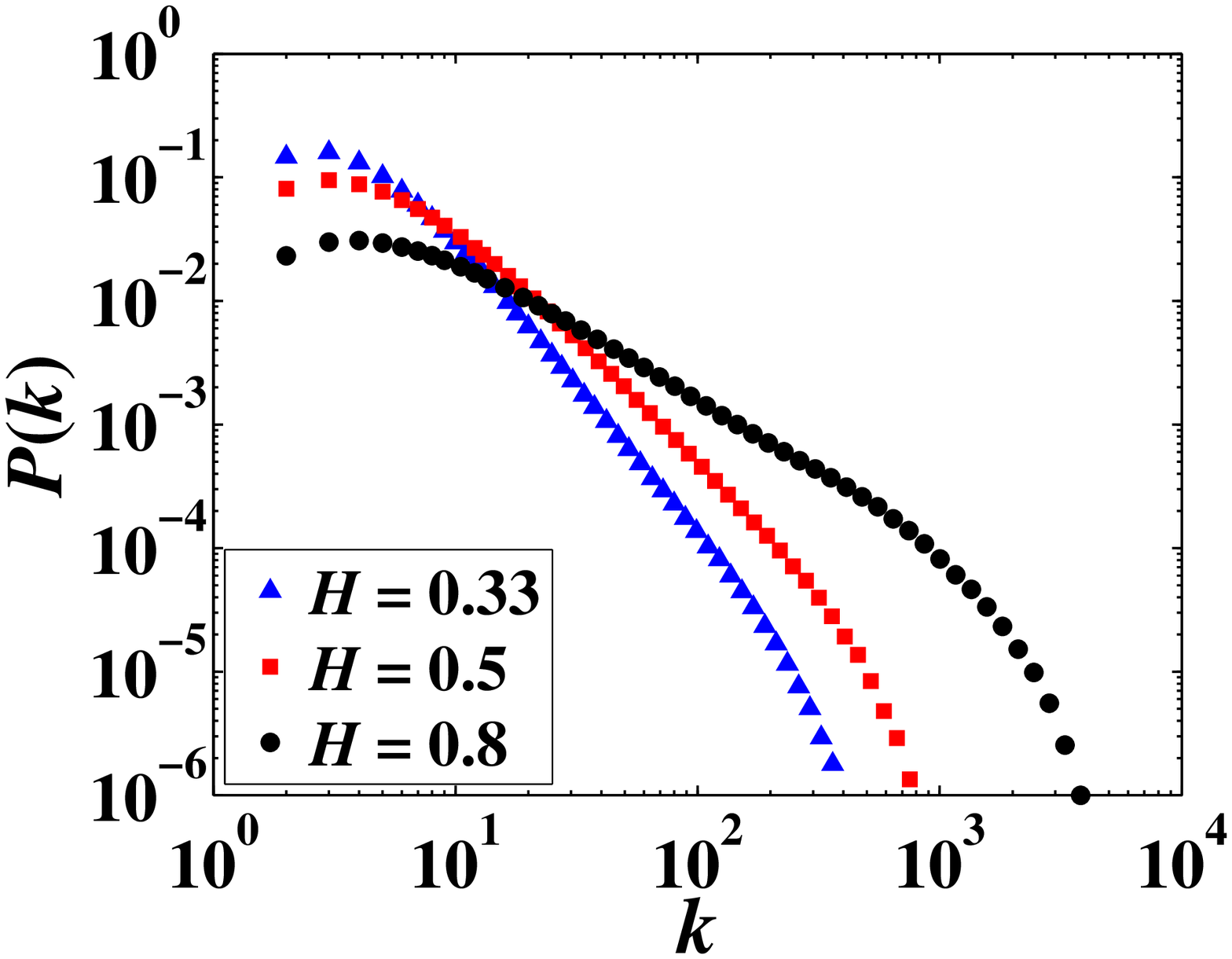}
\caption{(color online.) (a) Empirical degree distributions of the
visibility graphs converted from MRW series with different Hurst
indexes $H=0.33$, $0.5$ and $0.8$. (b) Empirical degree
distributions of the visibility graphs converted from FBM series
with different Hurst indexes $H=0.2$, $0.5$ and $0.8$.}
\label{Fig:FBM:MRW:pdf:k}
\end{figure}

We have investigated FBMs with the input Hurst index $H_{\rm{in}}$
ranging from 0.1 to 0.9 in the spacing of 0.1. For each
$H_{\rm{in}}$, we repeat the simulation 100 times and each
simulation gives a FBM signal with the size $N=50,000$. For each FBM
signal, a visibility graph is constructed and its empirical degree
distribution is determined. We find that the 100 distributions
almost collapse onto a single curve. This enables we to put all the
data of the 100 graphs to construct the empirical degree
distribution to gain better statistics. Three typical empirical
degree distributions of the visibility graphs converted from FBM
series with different Hurst indexes are depicted in
Fig.~\ref{Fig:FBM:MRW:pdf:k}(a). Nice power-law behaviors are
observed in the distributions, followed by faster relaxation. We
note that the visibility graphs of other FBMs also exhibit power-law
tails in the degree distribution.

The situation is very similar for the MRW case. We have simulated
MRW signals of size $N=50,000$ with different input Hurst index
$H_{\rm{in}}$ ranging from 0.05 to 0.95 with an increment of 0.05.
For each $H_{\rm{in}}$, 100 MRW signals are simulated and then
converted to 100 visibility graphs. Three typical empirical degree
distributions of the visibility graphs are depicted in
Fig.~\ref{Fig:FBM:MRW:pdf:k}(b) for different Hurst indexes $H$ (not
$H_{\rm{in}}$). All the distributions exhibits nice power laws with
faster decay for large degrees.

\begin{figure}[htb]
\centering
\includegraphics[width=7cm]{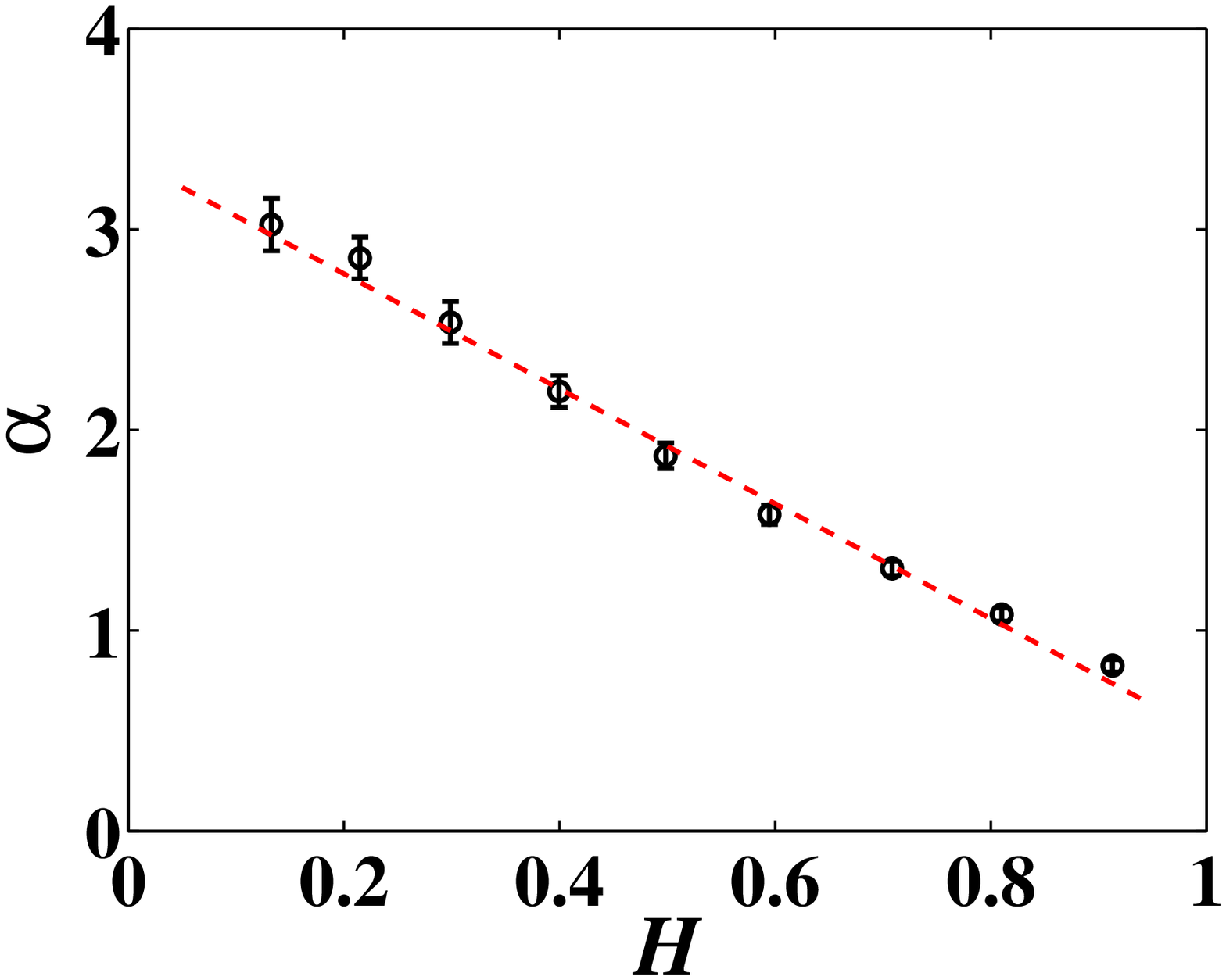}
\includegraphics[width=7cm]{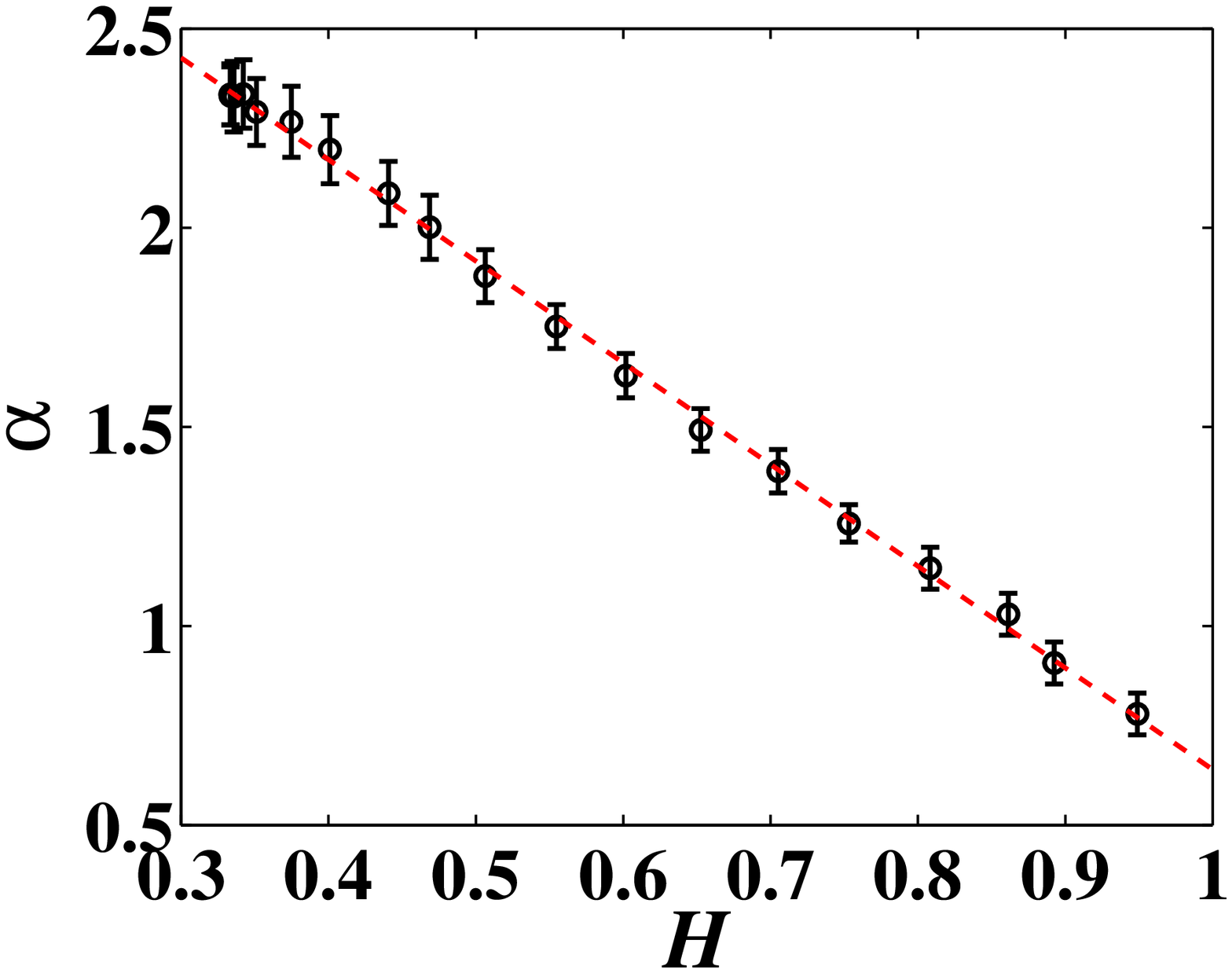}
\caption{(color online.) Dependence of the power-law exponent
$\alpha$ on the Hurst index $H$ for (a) FBM visibility graphs and
(b) MRW visibility graphs. The straight lines are the least-squares
fits of Eq.~(\ref{Eq:alpha:H}).} \label{Fig:alpha:H}
\end{figure}

The power-law exponents $\alpha$ of the distributions are calculated
in the scaling ranges. Figure \ref{Fig:alpha:H} shows the dependence
of the power-law exponents $\alpha$ on the Hurst indexes $H$. Both
curves show a nice linear relationship:
\begin{equation}
 \alpha(H) = a-bH~. \label{Eq:alpha:H}
\end{equation}
A least-squares regression gives $a=3.35$ and $b=2.87$ for FBMs and
$a=3.19$ and $b=2.55$ for MRWs. We find that the multifractal nature
of the MRWs has minor influence on the degree distributions of the
visibility graphs.

We note that, the linear relationship between the tail exponent and
the Hurst index was also found for fractional Brownian motions
independently \cite{Lacasa-Luque-Luque-Nuno-2009-XXX}.

\section{Applications to financial data}
\label{S1:Applications}

\begin{figure}[htb]
\centering
\includegraphics[width=7cm]{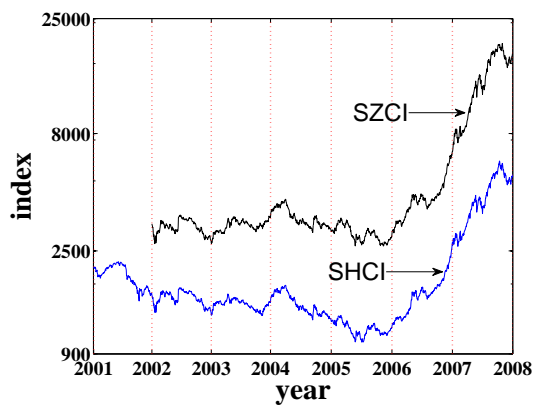}
\caption{Temporal evolution of the price trajectory of the Shanghai
Stock Exchange Composite index and the Shenzhen Stock Exchange
Component index.} \label{Fig:price}
\end{figure}

In this section, we apply the visibility graph method to financial
data. Specifically, two indices of the Chinese stock market are
considered. The organized stock market in mainland China is composed
of two stock exchanges, the Shanghai Stock Exchange (SHZE) and the
Shenzhen Stock Exchange (SZSE). Shanghai Stock Exchange Composite
index (SHCI) and Shenzhen Stock Exchange Component index (SZCI) are
the representative indices for the two stock exchange, respectively.
Our analysis is based on the 1-min data of the two indices. The time
series spans from 2 January 2001 to 28 December 2007 for the SHCI
and from 4 January 2002 to 28 December 2007 for the SZCI. The
temporal evolution of the price trajectories of the two indices are
illustrated in Fig.~\ref{Fig:price}. According to
Fig.~\ref{Fig:price}, the Chinese stock market was in a bearish
phase from 2001 to 2005, which is known as an antibubble
\cite{Zhou-Sornette-2004a-PA}, and then the market reversed and
produced a very marked bubble which burst at the end of 2007. In
order to test if the market phase has impact on the results, we
partition each time series into two sub-series delimited on 31
December 2005, corresponding to the bear period and the bull period,
respectively.

\begin{figure}[htb]
\centering
\includegraphics[width=7cm]{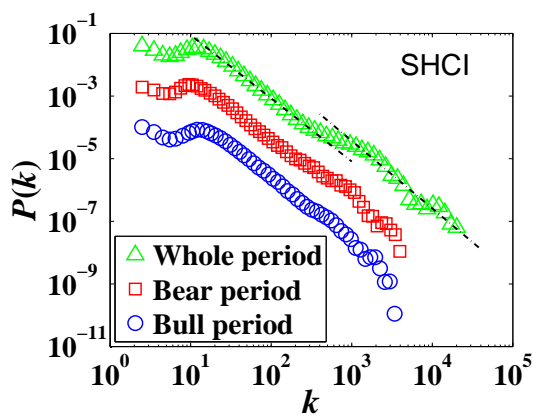}
\caption{(color online.) Empirical degree distributions of the three
visibility graphs of the Shanghai Stock Exchange Composite index.}
\label{Fig:pricedegreedist}
\end{figure}

For each index, we obtain three visibility graphs corresponding to
the bear period, the bull period, and the whole time period. The
empirical degree distributions of the three visibility graphs are
determined. Fig.~\ref{Fig:pricedegreedist} shows the results of
SHCI. We note that the results for the SZCI are very similar.

We observe that all the distributions have heavy tails and the
degree distribution in the bear period can be well modeled by a
power law (\ref{Eq:pk:PL}) in the right part. The rapid decay does
not mean that the power law is truncated but rather reflects the
fluctuations at finite size
\cite{Maillart-Sornette-Spaeth-vonKrogh-2008-PRL}. The tail exponent
is estimated to be $\alpha=1.95$. For the case of the bull period,
there is a hump in the tail. This phenomenon is more evident for the
whole period, which is caused by the fact that there are more points
in the bull period that can see more previous points (large
degrees). There are two power laws in the degree distributions for
the whole time period case, illustrated by two parallel dashed lines
in Fig.~\ref{Fig:pricedegreedist}. The estimates of the tail
exponents $\alpha$ for SHCI and SZCI are listed in Table
\ref{Tb:alpha:H}.

\begin{table}[htb]
 \centering
 \caption{Comparison of the estimated tail exponents $\alpha$ and the ``predicted''
          tail exponents $\alpha'$. The last row is $e=(\alpha-\alpha')/\alpha'$.}
\medskip
\label{Tb:alpha:H}
 \centering
 \begin{tabular}{ccccccccc}
   \hline \hline
  \multirow{3}*[1.5mm]{} & \multicolumn{3}{c}{SHCI}&&\multicolumn{3}{c}{SZCI}\\  %
  \cline{2-4}  \cline{6-8}
         & Bear & Bull & Whole && Bear & Bull & Whole  \\
    \hline
    $H$       & 0.52 & 0.50 & 0.51 && 0.54 & 0.51 & 0.52  \\
    $\alpha$  & 1.95 & 1.88 & 2.00 && 1.92 & 1.67 & 1.72  \\
    $\alpha'$ & 1.85 & 1.92 & 1.87 && 1.82 & 1.88 & 1.85  \\
    $e$       & 0.05 &-0.02 & 0.06 && 0.05 &-0.11 &-0.07  \\

    \hline\hline
  \end{tabular}
\end{table}

\begin{figure}[htb]
\centering
\includegraphics[width=7cm]{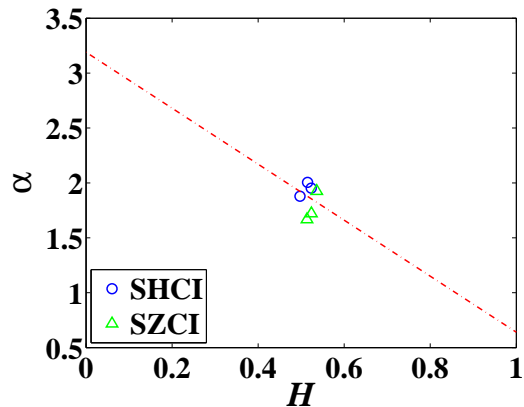}
\caption{(color online.) Scatter plot of $(H,\alpha)$ of the six
time series comparing the empirical results with the numerical
results predicted by $\alpha' = 3.19-2.55H$ (dot-dashed line).}
\label{Fig:hurstslope}
\end{figure}

In order to compare the empirical results with the numerical results
in Section \ref{S1:Numerical}, we determine the Hurst indexes of the
log returns for each index by applying the detrended fluctuation
analysis. The results are listed in Table \ref{Tb:alpha:H}. One
finds that the Hurst indexes are undistinguishable from $H=0.5$. The
multifractal detrended fluctuation analysis
\cite{Kantelhardt-Zschiegner-Bunde-Havlin-Bunde-Stanley-2002-PA}
confirms that all the time series possess multifractal nature.
Therefore, the tail exponents can be ``predicted'' according to the
linear relationship $\alpha' = 3.19-2.55H$. The predicted tail
exponents are also presented in Table \ref{Tb:alpha:H}. It is found
that the discrepancy between $\alpha$ and $\alpha'$ is not large,
which is quantified by the relative difference
$e=(\alpha-\alpha')/\alpha'$ shown in Table \ref{Tb:alpha:H}.
Fig.~\ref{Fig:hurstslope} further illustrate this point by the
scatter plot of the six data points $(H,\alpha)$. These data points
are very close to the dot-dashed line $\alpha' = 3.19-2.55H$,
showing that the empirical results are consistent with the numerical
analysis in Section \ref{S1:Numerical}.

\section{Conclusion}
\label{S1:Conclusion}

In summary, we have studied the degree distributions of visibility
graphs extracted from fractional Brownian motions and multifractal
random walks. We found that the degree distributions exhibit
power-law behaviors, in which the power-law exponent is a linear
function of the Hurst index inherited in the time series. In
addition, the degree distribution of the visibility graph is mainly
determined by the temporal correlation of the corresponding time
series, and contains minor information about the multifractal nature
of the time series. The linear relation (\ref{Eq:alpha:H}) provide a
possible tools for the determination of $H$ of a time series from
its visibility graph
\cite{Lacasa-Luque-Ballesteros-Luque-Nuno-2008-PNAS}. However,
cations should be taken since the increments distribution of the
time series might also have impact on the degree distribution.

\bigskip
{\textbf{Acknowledgments:}}

This work was partly supported by the National Natural Science
Foundation of China (70501011), the Program for New Century
Excellent Talents in University (NCET-07-0288), the Shanghai
Educational Development Foundation (2008SG29), and the China
Scholarship Council (2008674001).

\bibliography{E:/Papers/Auxiliary/Bibliography}

\end{document}